# Metal-less Metamaterial for Surface Plasmon Polariton guiding and amplification

Pavel Ginzburg and Meir Orenstein

*Micro-Photonics Lab., EE department, Technion – Israel Institute of Technology*
*Technion City, Haifa 32000 Israel*
*meiro@ee.technion.ac.il*

**Abstract:** We propose a novel metamaterial for Surface Plasmon Polariton guiding, amplification and modulation. Specific example of AlN/GaN Quantum Cascade Amplifier and its dispersion engineering are studied in details. The general original concept of metamaterials based on inclusions of low-dimensional quantum structures (artificial atoms) is discussed.

## Introduction

Nano-sized photonics, being the aim of interest of both fundamental theoretical concepts and practical implementations, gained a momentum in recent years, due to newly developed nano-fabrication techniques. While the dielectric-based optics does not allow the realization of nano-scaled optical devices, metal-based Surface Plasmon Polariton (SPP) seemingly does. SPP optical waves on a metal-dielectric interface can be confined much below the optical wavelength [2,3 and 4], and even offer a route to overcome the diffraction-limit [5]. Research of sub-wavelength SPP optics shows promise for the realization of nanometer-size Photonic Integrated Circuits (PIC) [2 and 9], for applications such as optical interconnects, signal processing and nano-sensing. Nanometeric metal stripes [6, 10 and 7], v-grooves [8], trances [11] and spheres [12] surrounded by a dielectric medium, can potentially serve as SPP waveguide devices, as was shown theoretically and experimentally.

The basic SPP phenomenon is the waveguiding on a single dielectric/metal interface [13], and the condition for the guiding is the different sign of electric permeability of the interfacing media (and relative magnitudes). While the common dielectrics exhibit positive electromagnetic response to the applied field, negative electric permeability is exhibited by noble metals in optical frequencies due to the plasma like behavior of conducting electrons. One of the most disappointing features of SPP, limiting their breakthrough into practical realizations, is the inherent losses, caused by the metal free carrier absorption, resulting in wave propagation distance in the micrometer scale [6]. Mitigating the loss problem by adding gain medium like electrically-pumped semiconductor bulk material or semiconductor quantum wells (QW) [14 and 15] is still not realized due to the very high gain requirement. In addition, the SPP waves are naturally highly confined in the very small region of the interface; making this kind of amplification procedure to be even more difficult (only a small portion of electromagnetic field overlaps the active layer). In summary, the inherent losses and lack of simple amplification solutions are significant barriers of nowadays existing configurations for the realization of SPP based nanometer-size PIC.

Since the ultimate material for SPP guiding does not exists in nature, the probable solution for this problem will be to engineer the artificial material with prescribed characteristics of SPP guiding. The concept of metamaterial was introduced some years ago [16], and more recently in [17]. The major effort was done towards realization of Left-Handed Material (LHM) [18 and 19], but other prescribed properties may be constructed as well [20].

Here we propose and analyze the concepts of metal-less metamaterial for SPP guiding in the optical regime, based on low dimensionality semiconductor quantum structures. The realization of plasmonics using semiconductor may revolutionize this field by allowing high level integration, dynamic plasmonic elements and the possibility of plasmonic amplification by carrier injection – all these features are expected to be the outcome of this single metamaterial.

**Basic Concept**

The Lorentz model of atomic electrical permeability is well-approved by advanced quantum theory and numerous experiments [21]:

$$\varepsilon = \varepsilon_{bac} - \sum_{i<j} iNq \frac{|\mu_{ij}|^2 (\rho_{ii} - \rho_{jj})}{\varepsilon_0 \hbar} \frac{1}{i(\omega - \omega_{ji}) - \gamma_{ji}} \Gamma \qquad (1)$$

where q is the electron charge, $\varepsilon_0$ is the vacuum electric permeability, $\omega_{ji}$ is the resonant transition frequency, $\omega$ is the central frequency of the input light, $\rho_{ii}$ is the electron occupation density of the i-th level, N is the carrier density participating in the process, $\varepsilon_{bac}$ is the averaged permeability of the background material, $\gamma_{ji}$ is phenomenology introduced dephasing rate and $\Gamma$ is the optical field confinement factor within the material. By employing the appropriate parameters in Eq. 1, it is possible to attain a negative real part of $\varepsilon$ in the vicinity of a transition resonance. Since the basic mechanism of SPP is presence of a medium with negative electrical permeability, the material, contains such kind of resonances may support the SPP guiding. Moreover, being also very lossy, the passive material (Fig. 1(a)) may be inverted to be active (Fig. 1(b)) by population inversion achieved by external pumping.

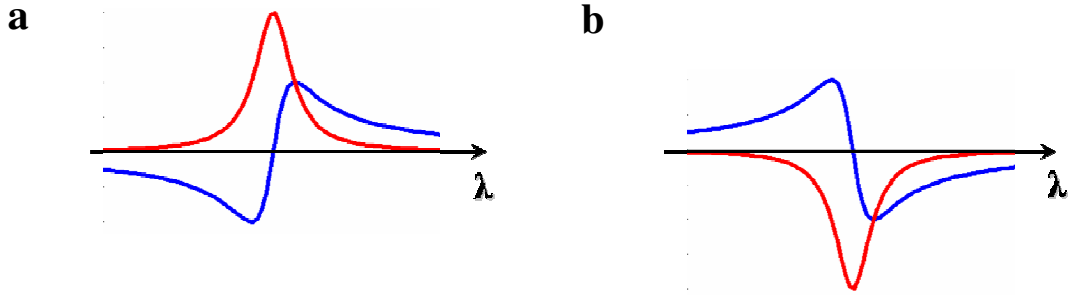

Fig. 1. Lorenz model of two-level atom permeability – imaginary part (red curve) and real part (blue curve); (a) electron in a ground state – passive configuration; (b) electron in an exited state – active configuration

A favorable candidate for a pronounced optical transition resonance is the semiconductor quantum well (QW). We will use QW structures to demonstrate the phenomenon due to their widespread usage and technology maturity; nevertheless, lower dimensionality structures such as quantum wires and quantum dots (QD) may exhibit similar performance.

An interesting point of view is to relate QW and QD to artificial atoms [22], since their well-approved growing techniques [23 and 24] is promising to achieve on demand fine structure of atomic levels. The outstanding concept to compose materials from such kind artificial atoms has been shown in our recent work [25].

As it was discussed, the existance of transition resonance it crucial for the reduction of background material permiability towards the negative values. Thus, the resonance linewidth plays the key role in the pemiability variation (Eq. 1). Interband transitions (Fig. 2(a)) are characterized by a wide absorption (emission) spectrum (unfortunate for our application), due to the inverse parabolicity of the conduction and valence bands. This does not allow for the related real permeability to exhibit the distinct negative region which is typical for narrow resonances. However, the intersubband transitions (Fig. 2(b)), having the same parabolicity for all energy levels, allow much narrower spectrum of interaction. The accuracy of this statement depends on k-dependence of effective mass model [21].

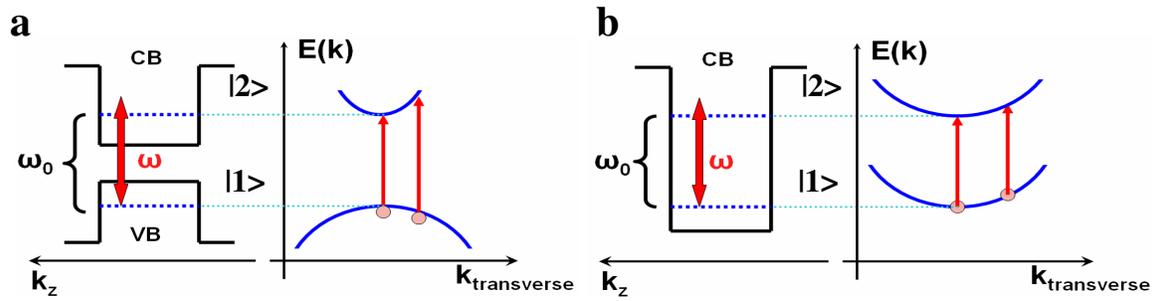

Fig. 2. Energy-level diagram quantum wells; (a) Interband transitions; (b) Intersubband transitions

Amplification via intersubband transitions may be achived by different schemes, as it was shown both theoretically and experementaly [26 - 33]. The most promising technique is Quantum Cascade (QC) [26-30]. Different material familes was sucsesfully used for QCL composition, but for the most of common semiconductors which are in use today, the resonance wavelengths for the intersubband transitions is in the far infrared region making them inapplicable for the optical range SPP applications. Fortunately, the Nitride group of materials, being the source for the blue lasers, does exhibit optical frequencies by intersubband transitions [34 and 35].

Nitride materials got a considerable attention during the recent years because of their special advantages, which will be discussed, and number of promising detectors [36] and emitters [37] were demonstrated recently.

## Negative ε within GaN/AlN Quantum Cascade Amplifier (QCA) at 1.55 μm

QCL generally constructed from a chain (cascade) of basic cells, each comprised of the active and transport regions. The basic concept of QCL amplification may be described by electrical carrier injection into the active region of the first cell where the radiative recombination takes place and the carrier subsequently is transported to the next cascade. In our application the transport region should be reduced in dimensions in order to maximize the confinement factor $\Gamma$ (Eq. 1). In this sense, tunnelling and scattering lateral transport techniques [28] are preferable over the miniband transport [26].

We use the idea of LO-phonon scattering transport in QCL [28] further presented in Nitrides [38]. The advantage of using Nitrides, accept of the operation wavelength in our application, mentioned in [38], are high energy LO-phonons (~90meV), allowing the room temperature operation of QCL, due to reduced parasitic thermal repopulation. Additional benefits of Nitrides are the low background dielectric constant [39], which should be reduced to the negative values by the fine quantum structures resonances, and possibility of high-speed control [40].

We modeled QCA in GaN/AlN by finite element simulation of 2nm AlN -2.45nm GaN -1.4nm AlN - 1.25nm cascade with 50 mV/nm applied voltage got the following conduction band diagram with the respective electrons' wavefunctions (Fig. 3)

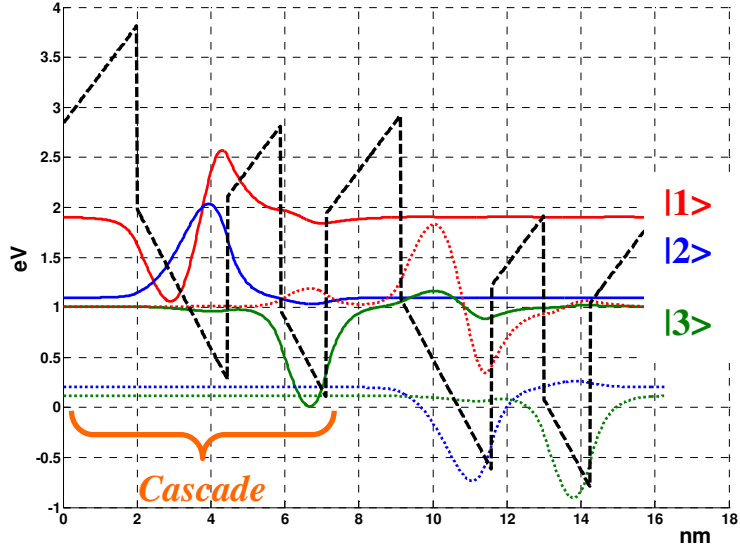

Fig. 3. Conduction band profile (black dashed line) of GaN/AlN QCA. Color solid lines represents the electron wavefunctions of the left cascade, while the color doted lines are the wavefunctions of the right cascade.

Being injected into left cascade, the electron occupies the exited level (|1>) of the QW – solid red line. The quantum efficiency of such injection process discussed in [22] and it is about a unity, for working devices. Making the radiative transition into the ground state (|2>) - blue solid line (by amplification) electron is resonantly scattered by LO-phonon [28] into the second well of the cascade (|3>) – green solid line. The applied bias voltage locate the lower state of the left cascade in resonance with the upper state of the next, allowing the electron tunnel efficiently to the next stage. Critical bottleneck of intersubband population inversion is a small lifetime of the exited states carriers. The influence of phonon scattering, electron-electron scattering, roughness scattering and spontaneous emission on carrier dynamics have been studied in literature [41 and 46]. As it was shown, LO-phonons are the most significant contributors to the carriers' lifetime. We used the formalism for lifetime calculations, presented in [28 and 41] and applied it on our configuration. We took into account only the GaN bulk modes, neglecting the contribution of the heterostructure. This approximation is widely used and gives good results [38], but more compound GaN/AlN phonon modes may be used as well [42]. The material parameters were taken from measured data – [43-45].

The corresponding electrical permeability of this QCA for charge sheet density $\rho_s = 2 \cdot 10^{13}$ cm$^{-2}$ is represented on Fig. 4. This density of injected carriers applicable and allows the voltage drop on the device.

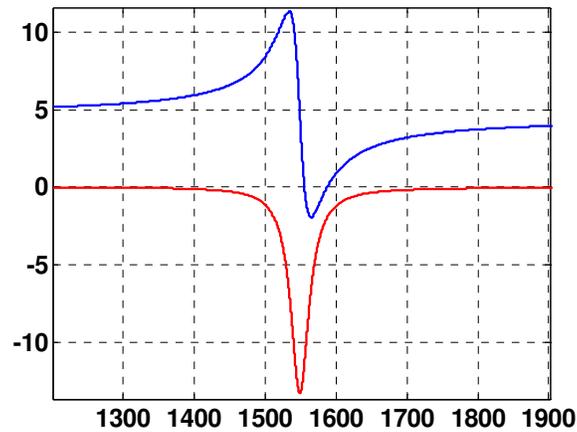

Fig. 4. Electrical permeability of GaN/AlN QCA. Blue line – real part; Red line – imaginary part

From Fig. 4 one may see, that the negative ε of -1.9-5.7j may be achieved around 1567nm central wavelength.

We got negative-ε material response of GaN/AlN QCA. The question now, if it is possible to improve somehow the performance of the device, at least in the meaning of the injected current, and achieve negative-ε response over the gain peak, significantly reducing the amplifier noise at the redundant frequencies ? (Eq. 1 evidently shows that in single resonance material is not possible to achieve variation of real part of permeability at the gain/absorption peak).

## Advanced dispersion engeneeging

We propose to include an absorption peak in the vicinity of gain line to achieve enhanced ε response, especially to better coincide the maximum gain and negative real part of ε at the same wavelength. Consider two situations, depicted on Fig. 5:

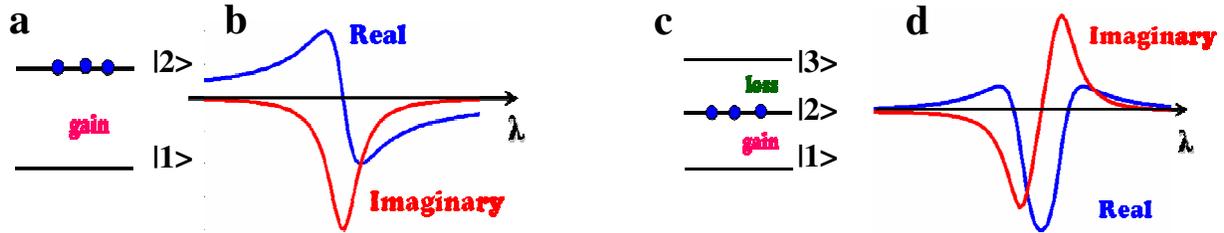

Fig. 5. Advanced engineering scheme – (a) two-level system with inverted population; (b) dispersion curves for system in "a"; (c) tree-level system – |2>-|1> transition corresponds to gain, while |2>-|3> transition corresponds to loss; (d) dispersion curves for system in "c"

By appropriate construction of this "three-level" system, one may see from Fig.5 (d), that the requested performance indeed achieved.

## Negative ε within Advanced QCA

Considering the discussed material family, performing a similar finite elements simulation we got the following structure – Fig. 6. The corresponding lifetimes of LO-scattering process are ~0.1psec for |1> and ~2.6psec for |2>, and this improves the possibility of efficient achievement of intersubband population inversion. The voltage drop on the device is 0.17 V/nm.

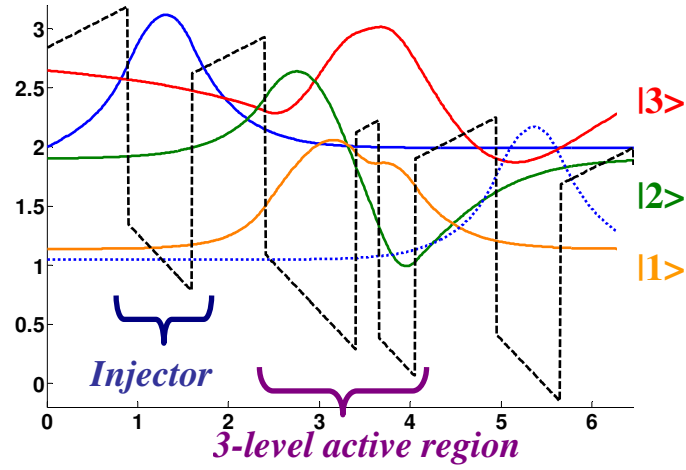

Fig. 6. Conduction band profile (black dashed line) of GaN/AlN QCA. Color solid lines represents the electron wavefunctions of the left cascade, while the color doted line is the wavefunction of the right cascade injector level.  Orange, green and red represents the levels on Fig. 5(c).

The resulting electrical susceptibility of this QCA for different injection levels is depicted on Fig. 7:

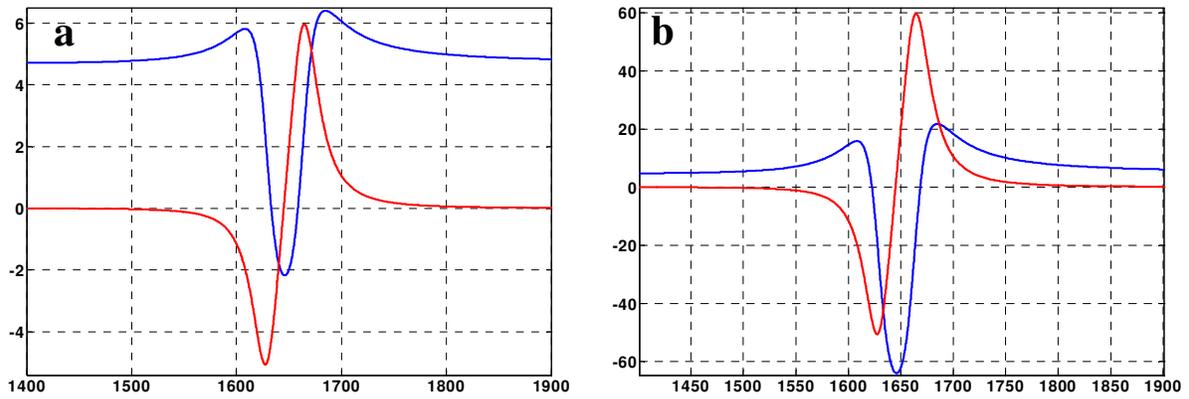

Fig. 7. Electrical permeability of GaN/AlN QCA. Blue line – real part; Red line – imaginary part. (a) $\rho_s=5 \cdot 10^{12}$ cm$^{-2}$ (b) $\rho_s =5 \cdot 10^{13}$ cm$^{-2}$; $\rho_s$ is the sheet density of injected carriers

From Fig. 7 one may see, that that negative $\varepsilon$ may be achieved: for the sheet density of injected carriers $\rho_s=5 \cdot 10^{12}$ cm$^{-2}$ -2-2j around 1640 nm central wavelength and for $\rho_s=5 \cdot 10^{13}$ cm$^{-2}$ -63-2.3j (1646 nm – lowest real value) and -24-50j (1629 nm – maximum gain).

**Conclusions**

We discussed the possibility to achieve negative electrical permeability within the metamaterial, based on QCA structure. This single metamaterial contains within the crucial properties of guiding, amplification and modulation of SPP that are the cornerstone of any future optical plasmon-based chip, to be build. We also introduce a novel concept of metamaterial construction from artificial atoms, like QW and QD. However – it should be remembered that constructing GaN based QCA is currently a formidable task – significant improvement of interfaces is required for exhibiting efficient tunneling, and the gain required for our scheme is extremely high. Yet this material maturity (due to other applications of the Nitride family) is expected to yield a significant impact in the metamaterials for plasmonics.